\documentclass[twoside,twocolumn,english,aps,prb,10pt]{revtex4}
\usepackage{natbib}[2009/07/16]
\usepackage[T1]{fontenc}
\usepackage[utf8]{inputenc}
\usepackage{textcomp}
\usepackage{amstext}
\usepackage{amsmath}
\usepackage{graphicx}
\usepackage{babel}

\begin{document}

\title{Influence of elastic scattering on the measurement of core-level
binding energy dispersion in x-ray photoemission spectroscopy}

\author{E. F. Schwier$^{1}$, C. Monney$^{2}$, N. Mariotti$^{1}$, Z. Vydrov\`a$^{1}$,
M. Garc\'ia-Fern\'andez$^{1}$, C. Didiot$^{1}$, M. G. Garnier$^{1}$,
P. Aebi$^{1}$}

\affiliation{$^{1}$\textit{\footnotesize D\'epartement de Physique and Fribourg
Center for Nanomaterials, Universit\'e de Fribourg, CH-1700 Fribourg,
Switzerland}}

\affiliation{$^{2}$\textit{\footnotesize Research Department Synchrotron Radiation
and Nanotechnology, Paul Scherrer Institute, CH-5232 Villigen PSI,
Switzerland}}

\date{\today}

\begin{abstract}
In the light of recent measurements of the C 1s core level dispersion
in graphene {[}Nat. Phys. \textbf{6}, 345 (2010){]}, we explore the
interplay between the elastic scattering of photoelectrons and the
surface core level shifts with regard to the determination of core
level binding energies in Au(111) and Cu$_{3}$Au(100). We find that
an artificial shift is created in the binding energies of the Au 4f
core levels, that exhibits a dependence on the emission angle, as
well as on the spectral intensity of the core level emission itself.
Using a simple model, we are able to reproduce the angular dependence
of the shift and relate it to the anisotropy in the electron emission
from the bulk layers. Our results demonstrate that interpretation
of variation of the binding energy of core-levels should be conducted
with great care and must take into account the possible influence
of artificial shifts induced by elastic scattering.
\end{abstract}
\maketitle

\section{introduction}

The general assumption of non dispersing core levels is only valid
for fully localized states, but due to the continuous nature of the
electronic wave function, the orbitals of core levels can exhibit
weak hybridization even for binding energies of up to a few hundred
eV. This has been demonstrated for gaseous molecules like C$_{2}$H$_{2}$
\citep{Kempgens:1997p3098} and N$_{2}$ \citep{Hergenhahn:2001p3092}
and in recent measurements on graphene by Lizzit \emph{et al.} \citep{Lizzit:2010p562}
who were able to determine the bandwidth of the C 1s core level to
be 60 meV. The corresponding theoretical prediction from \emph{ab
initio} calculations support the claim, that these orbitals are not
completely degenerated.

To resolve the dispersion of core levels and the size of the Brillouin
zone in solids with x-ray photoelectron spectroscopy (XPS), high angular
and energy resolutions are necessary. In addition to that, there exist
two mechanisms that introduce a broadening into the measurement. These
are the quasi elastic scattering of the photoelectron with the atoms
of the crystal \citep{Boersch:1967p890,Erickson:1989p49} as well
as the influence of phonons on the photoemission process \citep{White:1986p1940,Plucinski:2008p298}.
A third factor which has not been considered until now and that can
introduce a systematic error to the measurement of the dispersion
in core-level binding energies is introduced in this paper with the
demonstration of an angular dependent artificial shift, created by
the existence of unresolved energetically shifted surface core levels.

The surface core level shift ($\Delta E_{SCLS}$ ) describes the energy
shift between the core levels attributed to the bulk and the surface
of a crystal \citep{Houston:1973p1435}. It arises from a combination
of initial \citep{Johansson:1980p3099} and final state \citep{Pehlke:1993p3100}
effects. The non continuous charge distribution and the reduced number
of neighbouring atoms at the surface of the crystal can change the
coulomb potential in the topmost layers with respect to its bulk value.
This leads to a shift in the core level binding energies of the surface
atoms. In addition to this, interactions between the photoelectron
and a partially screened photohole, that is created during the photoemission
process, will also influence the surface core level shift. Measurements
on noble metal films \citep{CITRIN:1978p2114}, 5d metals \citep{Mrartensson:1989p611}
graphite \citep{Prince:2000p960} and rare-earth crystals \citep{Shamsutdinov:2005p2580}
have shown that the displacement between bulk and surface core levels
can be positive as well as negative and possesses an amplitude of
up to several hundred meV.

\begin{figure}[h]
\includegraphics{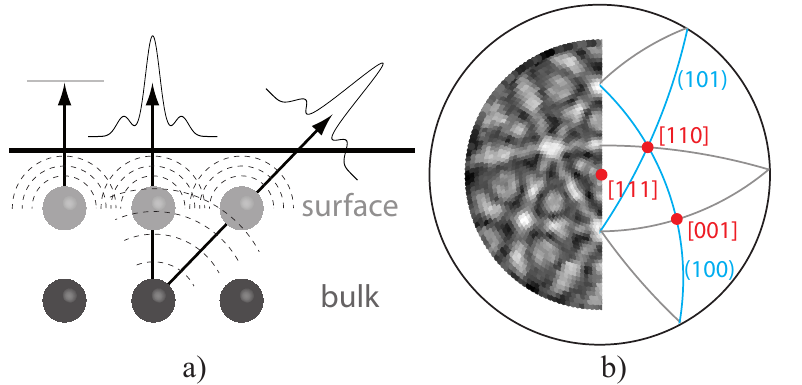}

\caption{a) Contribution to the scattering amplitude from the surface atoms
(light grey) and the bulk atoms (dark grey). Closed packed directions
in the bulk change the scattering amplitude for different emission
angles. The contribution from the surface layer remains almost constant
up to grazing emission. b) A typical 2 $\pi$ angular scan in stereographic
projection with intensity anisotropy due to scattering of photoelectrons.
The high symmetry directions are visible as forward focusing peaks
(points), while high symmetry planes in the bulk manifest themselves
as Kikuchi bands (lines). Normal emission is plotted at the center
while the black circle corresponds to emission at grazing angles.
Maximum and minimum intensity corresponds to white and black, respectively.\label{introXPD}}

\end{figure}

It is important to note that at kinetic energies above 500 eV the
strong anisotropy in the individual electron-atom scattering leads
to a focusing of electron flux along directions pointing from the
emitting atom to the scatterer (Fig. \ref{introXPD} a) \citep{Siegbahn:1970p1716}.
This effect is also known as x-ray photoelectron diffraction (XPD).
In 2$\pi$ angular XPD patterns (Fig. \ref{introXPD} b) high symmetry
directions as well as low index lattice planes can be identified through
forward focusing peaks and Kikuchi lines. Both structures are used
to provide local information about the atomic structure near the surface
\citep{Fadley:1987p2206,Chambers:1991p3003,Fasel:1995p2288}.

In this work we focus on the influence of the elastic scattering of
photoelectrons on the binding energy of core levels. We demonstrate
a correlation between the binding energy and the emission intensity
of the core level and propose a mechanism that explains the angular
dependence of core level binding energies.

\section{Experiment}

All measurements were performed with an upgraded SCIENTA SES 200 analyzer,
allowing for multiple angle parallel detection, using a non-monochromatized
MgK$_{\alpha}$ ($h\nu=1253.6\: eV$) x-ray anode as excitation source.
A computer controlled 5-axis manipulator allows rotations of the sample
along the polar and azimuthal directions with a precision of 0.1\textdegree{}
and 0.2\textdegree{}, respectively. The angular acceptance of the
entrance hole was 2.4\textdegree{}. The energy steps of the spectra
were set to 19 meV for the 2$\pi$ angular scans and to 50 meV for
the other measurements. All spectra were taken at room temperature
with parallel detection in angle and energy. During the measurement,
the pressure in the chamber did not exceed 2$\times$10$^{-10}$ mbar.
The crystals were prepared with multiple sputter and annealing cycles.
The sputtering acceleration voltage was set to 1.5 kV and the incident
angle of the argon ions was chosen to be 65\textdegree{} off normal
while the sample was rotated.

To determine the influence of the elastic scattering of photoelectrons
on the binding energy of core levels, we have chosen the Au 4f doublet
of the Au(111) and Cu$_{3}$Au(100) surfaces, as they exhibit a relatively
large surface core-level shift (Au(111): $\Delta E_{SCLS}=0.35\pm0.01\: eV$
\citep{HEIMANN:1981p1563} and in the case of Cu$_{3}$Au(100): $\Delta E_{SCLS}=0.5\pm0.05\: eV$
from the works of DiCenzo et \emph{al.} \citep{DiCenzo:1986p1848}
and $\Delta E_{SCLS}=0.41\pm0.01\: eV$ \citep{Eberhardt:1985p453}).
Also, the surface preparation and properties of the Au(111) herringbone
reconstruction \citep{Barth:1990p1553,Reinert:2004p2201} as well
as the Cu$_{3}$Au(100)-c(2x2) surface reconstruction \citep{Buck:1983p645,STUCK:1990p2944,Dosch:1991p1212}
are well documented. 

The annealing temperature of the Au crystal was set to 600 \textdegree{}C
in order to obtain the herringbone surface reconstruction. The Cu$_{3}$Au
was heated up to 450 \textdegree{}C and cooled down within several
hours across the transition temperature (T$_{C}=390$ \textdegree{}C)
to obtain the c(2x2) reconstruction. The pressure never exceeded 8$\times$10$^{-10}$
mbar. After each preparation cycle, the surface order and cleanliness
were tested with LEED and XPS measurements, respectively.

\section{Results}

\begin{figure}
\begin{centering}
\includegraphics[width=0.9\columnwidth]{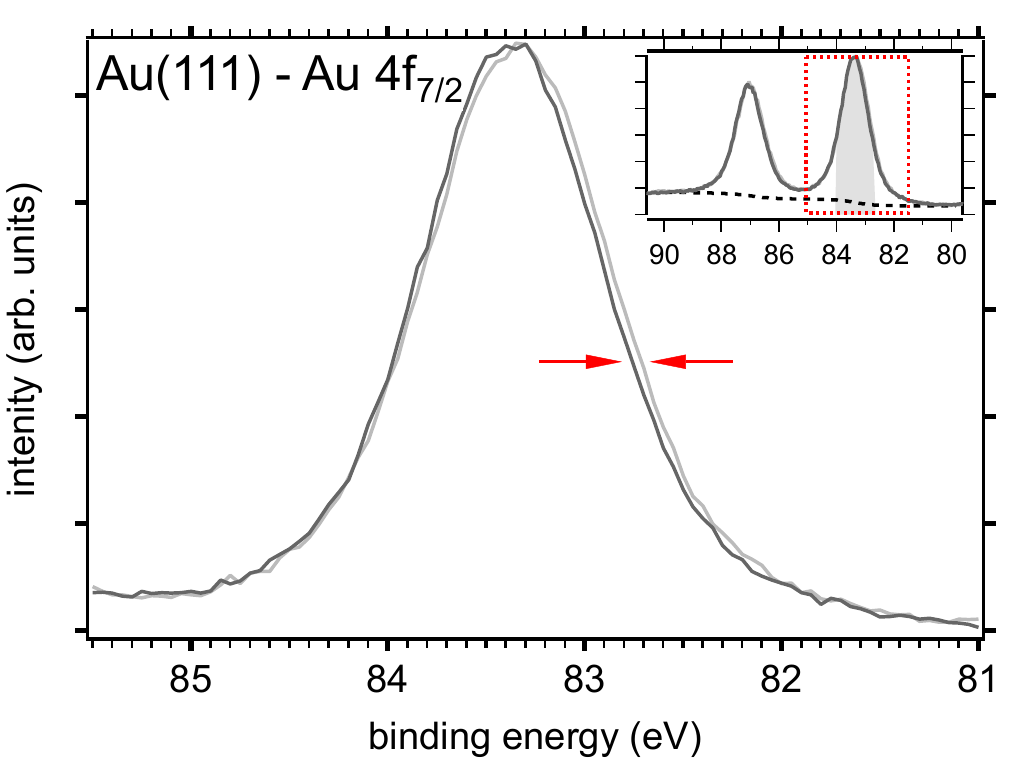}
\par\end{centering}

\centering{}\caption{Two spectra of the Au 4f$_{7/2}$ core level of Au(111), taken at
68\textdegree{} off normal at azimuthal angles corresponding to the
{[}001{]} symmetry direction (dark grey) and 12\textdegree{} off symmetry
(light grey). A shift of 40 meV (arrows) is visible. Both spectra
are normalized to the maximum intensity of the Au 4f$_{7/2}$ peak.
Inset: The whole Au 4f doublet with Shirley background. The grey area
below the peak corresponds to the interval used for the general fitting
procedures. (see text)\label{XPS_BvsS}}

\end{figure}

Two spectra (Fig. \ref{XPS_BvsS}) of the Au 4f doublet on Au(111)
were taken at the same polar angle, but at two different azimuthal
angles. While one was measured in a direction that coincides with
a low index crystal direction, the other spectrum was measured 12\textdegree{}
away from that symmetry point. The energy broadening of the Au 4f
peak allows to approximate its shape with a Gaussian profile including
a constant background. Fitting the two Au 4f$_{7/2}$ peaks with this
profile gives a shift of $\Delta E_{B}=40\pm5\: meV$ between the
binding energy of the two spectra. Considering the relatively high
electron energies and the broadening inherent to the XPS experiment
here, it is clear, that the shift cannot by attributed to a dispersion
induced by weakly hybridized Au 4f core levels.

To determine the angular dependence of the shift in the spectra of
Au(111), a 2$\pi$ solid angle emission scan of the Au 4f doublet
was measured. The intensity of the electron emission was recorded
as a function of the core level binding energy $E_{B}$, the polar
angle $\Theta$ and the azimuthal angle $\Phi$, generating a complete
set of energy spectra as a function of the emission angles. In addition
to that a polar angle was chosen, were a high resolution azimuthal
scan was performed. The resulting spectra $I\left(E_{B},\Theta,\Phi\right)$
were fitted in energy with a Gaussian profile, including a constant
background. To test the stability of the fit, an increased fitting
interval as well as a fixed width for the Gaussian profile were implemented
into the fit without changing the quality of the results. In addition,
the same behavior was found for the Au 4f$_{5/2}$ peak or upon using
a Voigt profile with a nonzero Lorenzian component or a Shirley type
\citep{Shirley:1972p3835} background.

\begin{figure}
\includegraphics{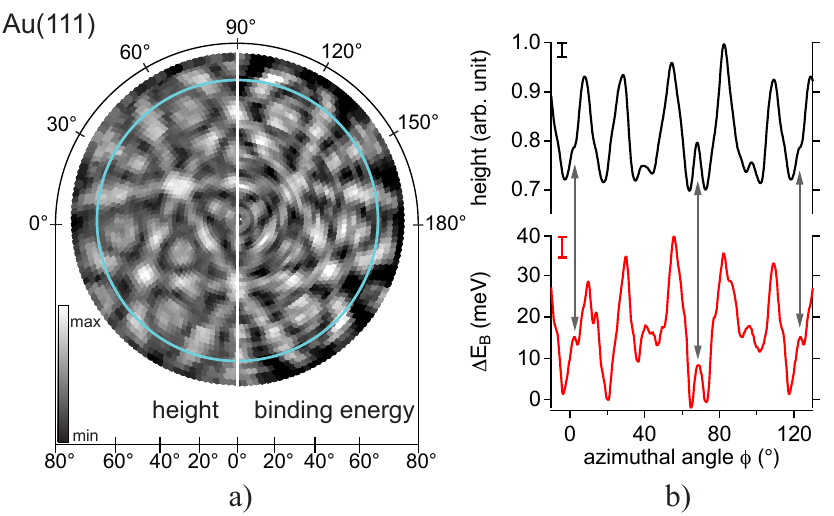}\caption{a) Stereographic projection of the angular dependence of the height
(left half) and binding energy (right half) of the Gaussian profile
used to fit the Au 4f$_{7/2}$ peak of Au(111).\emph{ }The position
of the azimuthal cut is marked in blue b) Plot of the height $h$
(black) and binding energy $E_{B}$ (red) of the Au 4f$_{7/2}$ peak
from Au(111) as a function of azimuthal angle, taken at $\Theta=68\text{\textdegree}$.
Grey arrows mark corresponding fine structures. The error bars from
the fitting parameters are in the top left of the azimuthal plots.\label{heightVScenter}}

\end{figure}

To increase the contrast of the anisotropy in the peak height $h$($\Theta,\Phi$)
and the binding energy shift $\Delta E_{B}$($\Theta,\Phi$) of the
Au 4f$_{7/2}$ peak a smooth background was subtracted from each dataset.
The resulting diffraction pattern are plotted in a stereographic plot
(Fig. \ref{heightVScenter} a). Comparison of the height and binding
energy shows that the main features in the scattering anisotropy,
i.e., the forward focusing peaks as well as the Kikuchi lines, are
clearly visible in both height and binding energy. Even fine structures
of the anisotropy are replicated in the binding energy as it can be
seen in the data from the high resolution azimuthal scan (Fig. \ref{heightVScenter}
b, arrows). As all of these features correspond to electron diffraction
induced by the bulk ordering of low index closed packed crystal directions
and planes a bulk mediated effect can be proposed. 

The binding energy of the Au 4f$_{7/2}$ exhibits a decrease of several
100 meV in the untreated data while changing the polar angle from
normal to grazing emission. This shift is proportional to $1/\cos\Theta$
and can be interpreted as an increased ratio between the emission
from surface and bulk core levels caused by the finite electron mean
free path. Even though this variation is only dependent on the polar
angle, it suggests that the surface core level shift plays a role
in the original observed shift shown in Fig. \ref{XPS_BvsS}. To support
this hypothesis, the Cu$_{3}$Au(100) alloy was chosen for further
measurement, as it exhibits a larger surface core level shift and
should therefore exhibit a larger amplitude in $\Delta E_{B}$ during
the azimuthal scans.

\begin{figure}
\includegraphics{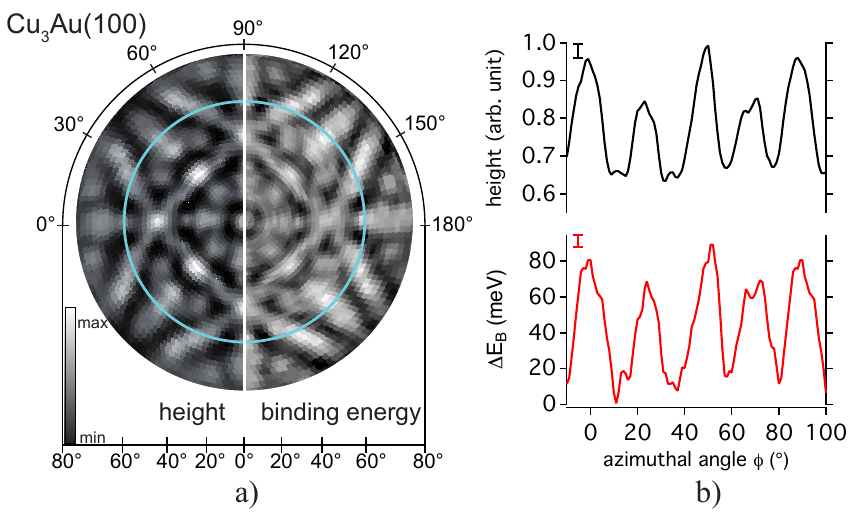}\caption{a) As Fig. \ref{heightVScenter} a) for Cu$_{3}$Au(100). b) As Fig.
\ref{heightVScenter} b) for Cu$_{3}$Au(100) with $\Theta=58\text{\textdegree}$.
\label{heightVScenter-1}}

\end{figure}

The same 2$\pi$ angular scan and a corresponding high resolution
azimuthal scan were performed for the Cu$_{3}$Au compound. The correlation
between the peak height $h$($\Theta,\Phi$) and the binding energy
$E_{B}$($\Theta,\Phi$) of the Au 4f$_{7/2}$ peak (Fig. \ref{heightVScenter-1}
a and b) is similar to the results found for Au(111). A comparison
between the maximum amplitude of the shift $\Delta E_{B}$ for Au(111)
($\Delta E_{B}\approx40\: meV$, Fig. \ref{heightVScenter} b) and
Cu$_{3}$Au(100) ($\Delta E_{B}\approx90\: meV$, Fig. \ref{heightVScenter-1}
b) supports the proposed influence of the size of the surface core
level shift on the amplitude of the angular dependence in the binding
energy shift. %

\section{Model}

\begin{figure}
\centering{}\includegraphics[width=0.9\columnwidth]{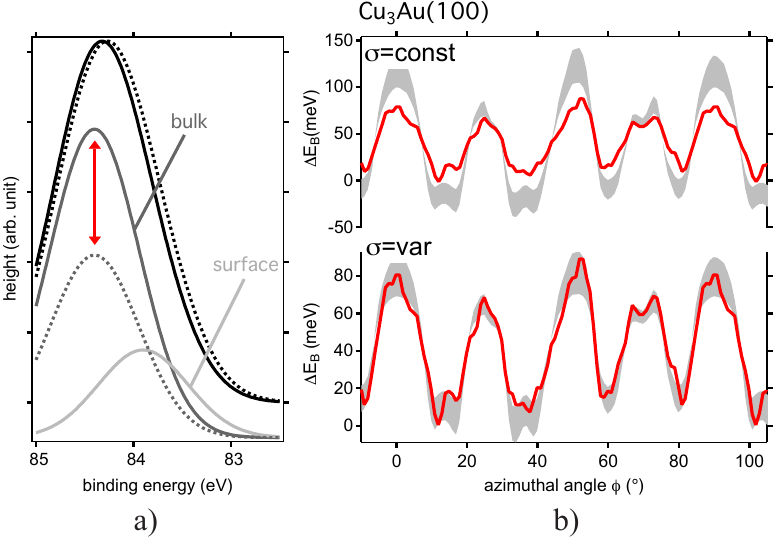}\caption{a) Peak composition considered for the model. The bulk peak (dark
grey) changes its height (dotted line) as a function of azimuthal
angle, while the surface contribution (light grey) stays constant.
The composite peak (black) exhibits an apparent shift (dotted line).
b) Comparison of the measured binding energy shift $\Delta E_{B}$
(red) with the confidence bands (see text) from a model peak composition
assuming a constant (grey, top) and variable (grey, bottom) bulk peak
width.\label{fitWithBroadening}}

\end{figure}

This dependence of the binding energy on the emission angle can be
understood if the different scattering amplitudes between photoelectrons
from the surface and bulk components are considered. We propose a
simple model (Fig. \ref{fitWithBroadening} a), that by considering
the Au 4f$_{7/2}$ peak as a sum of a surface and a bulk component,
can describe the measured azimuthal dependence of $\Delta E_{B}$.
The model peak consists of two Gaussian peaks with identical width
($\sigma=1.3\: eV$), corresponding to our experimental broadening.
We assume that the measured anisotropy in the azimuthal scan (Fig.
\ref{heightVScenter-1} b) is caused entirely by the angular dependence
of the scattering of the bulk component. The mean surface to bulk
ratio ($S/B=0.55$) for an emission angle of $\Theta=58\text{\textdegree}$
on Cu$_{3}$Au, was calculated from the results of DiCenzo \emph{et
al.} \citep{DiCenzo:1986p1848}. To account for the uncertainty of
the binding energy separation between the surface and the bulk component,
two sets of fits were performed using values for $\Delta E_{SCLS}$
corresponding to the upper ($\Delta E_{SCLS}=0.55\: eV$ \citep{DiCenzo:1986p1848})
and lower ($\Delta E_{SCLS}=0.40\: eV$ \citep{Eberhardt:1985p453})
limit of the literature values for the surface core level shift of
Cu$_{3}$Au. The variation in the fit between the two values is visualized
through a confidence band in the plot.

The resulting composite peak was fitted with a Gaussian profile, in
the same manner as the measured data and the binding energy shift
was plotted as a function of the azimuthal angle. In Fig. \ref{fitWithBroadening}
b (top) the dependence of the calculated binding energy (grey band)
is compared to the experimental results (red). The model predicts
a shift $\Delta E_{B}$, which is larger than the experimental results.
This can be corrected by relaxing the oversimplified assumption of
a Gaussian profile with a constant peak width. Note also that, using
a Voigt Profile instead of a Gaussian does not change the overestimation
of the binding energy shift. 

A good agreement between the model and the experiment can be achieved
if the width of the Gaussian peak corresponding to the bulk emission
is allowed to vary (Fig. \ref{fitWithBroadening} b, bottom). A possible
mechanism that would lead to a variation of the peak width is the
increase of the information depth through electron diffraction. Along
directions were forward focusing takes place, the electrons contributing
to a core level spectra will on average be emitted from a greater
depth and travel a longer path through the solid, thus increasing
the inelastic losses of the photo electrons. Another possible source
for a peak broadening is the change in cross section for scattering
events with the atoms in the closed packed directions, leading to
an increase in quasi elastic losses.

In a first order approximation, the broadening is assumed to scale
linearly with the intensity variation of the bulk peak. This can be
reasoned by comparing the information depth with the emitted intensity,
which both scale with the mean free path. The gain in the intensity
that is created through the forward focusing, will increase the mean
depth from where electrons are emitted and therefore increase the
possibility of occurrences of processes that involve energy loss.
This approximation is still applicable, if the defocusing of photoelectrons
through multiple scattering events in forward focusing directions
\citep{Tong:1985p832,Egelhoff:1987p445,Aebischer:1990p730} is taken
into account, as it effectively introduces a scaling factor to the
intensity emitted in forward focusing directions.

If a linear dependence of the bulk peak width on the bulk anisotropy
is implemented into the model ($\sigma(\Phi)=\sigma_{0}h(\Phi)/h_{0}$,
with $\sigma_{0}$ and $h_{0}$ being the average peak width and height,
respectively) and the resulting composite peak is again fitted in
the same way as described above, the calculated confidence band for
the binding energy shift (Fig. \ref{fitWithBroadening} b (bottom))
coincides with the results from the experiment. 

The amplitude of the shift $\Delta E_{B}$ is comparable to the predicted
and measured bandwidths, created through the hybridisation of partially
hybridized states \citep{Lizzit:2010p562}. In experiments that probe
the dispersion of such states, it is possible that an artificial shift
as described above, induces a systematic error. 

In the absence of an unresolved surface core level, a shift will nonetheless
be present if at least two energetically separated lattice sites for
the same element exist and if the coordination of those sites inside
the lattice leads to distinct electron scattering patterns. In that
case, the anisotropic electron emission and the correlation between
the intensity of the core level emission and its binding energy, would
be even more complicated.

Even in the case of measurements on systems as simple as monolayers,
the emission from two differently coordinated sites within the layer
(i.e., C 1s in graphene with partial H adsorption \citep{Balog:2010p3108})
should lead to a shift through multiple scattering within the layer
and with the substrate atoms. However, as the extend of the scattering
of such processes is usually weak compared to bulk mediated scattering,
the amplitude of such shifts is expected to be weaker as well.

\section{Conclusion}

We have demonstrated that the emission angle dependent shifts in the
binding energy of the Au 4f$_{7/2}$ core level, that were observed
on Au(111) and Cu$_{3}$Au(100) surfaces, can be explained by a simple
mechanism based on a peak composition that includes an energetically
unresolved surface core level. Our model uses the surface core level
shift, the intensity ratio between the surface and bulk emission,
the overall peak broadening and the diffraction of photoelectrons
emitted from bulk lattice sites. It is able to explain the shift in
binding energy between measurements at different emission angles and
its correlation to the emission intensity.

The measured shifts are qualitatively described by a model that assumes
a constant width for the bulk contribution. However, if a dependence
of the peak width to the anisotropy of the emission is introduced,
the experimental results and the calculated values match within the
error range.
\begin{acknowledgments}
This project was supported by the Fonds National Suisse pour la Recherche
Scientifique through Division II and the Swiss National Center of
Competence in Research MaNEP.

We would also like to thank the mechanical workshops in Neuch\^atel
and Fribourg for their support.
\end{acknowledgments}
\bibliographystyle{apsrev4-1}
\bibliography{paper}


\end{document}